\documentstyle[aps,prb,twocolumn]{revtex}
\begin{document}
\title{Tests of magnetic Hamiltonians for CaV$_4$O$_9$}
\author{Martin P. Gelfand}
\address{Department of Physics, Colorado State University, 
Fort Collins, Colorado 80523}
\author{Rajiv R. P. Singh}
\address{Department of Physics, University of California, 
Davis, California 95616}
\maketitle
\begin{abstract}
We study the uniform magnetic susceptibility and spin-gap for
recently proposed Heisenberg model Hamiltonians 
for CaV$_4$O$_9$ based on the orbital
ordering scenario of Marini and Khomskii and the LDA calculations
of Pickett. We argue that the experimentally 
observed uniform susceptibility data
is inconsistent with the weakly coupled dimer picture of Marini and Khomskii.
The model proposed by Pickett can, with appropriate choice of
parameters, lead to an explanation for the the observed gap and
uniform susceptibility. The resulting agreement with experiments 
is of a similar quality to previously
studied models. We argue that this new model is best distinguished
from previous ones by neutron or Raman scattering
experiments, via the location of the excitation minimum in the
Brillouin zone and by the possible existence 
or non-existence of sharply defined
singlet excitations.
\end{abstract}

\narrowtext

The observation of spin gaps in the calcium vanadates, especially in the
quasi-2D material CaV$_4$O$_9$, has attracted much attention. A number
of theoretical papers have explored the compatibility of the
experimental data with various model spin Hamiltonians. Ignoring some
out-of-plane distortions, a given layer of these materials forms a
one-fifth depleted square lattice with a four-spin unit cell 
(see Fig.~\ref{lattice}). 
It is by now well established that a uniformly nearest-neighbor coupled, 
one-fifth depleted square-lattice Heisenberg model has long range 
magnetic order.\cite{troyer,white,gwsoh,starykh} Thus the spin gap has
been argued to arise either (i) from non-uniform couplings which can lead
to weakly coupled plaquettes 
or (ii) from frustrating second-neighbor interactions,
which destabilize the N\'eel order. Either situation leads to a
``plaquette RVB'' (PRVB) ground state, and lies in the same $T=0$ phase
as a Hamiltonian which contains only intraplaquette nearest-neighbor
interactions.
Based on extensive comparison
of the experimental susceptibility and spin gap with theoretical
models, it was argued by Gelfand {\it et al.}\cite{gwsoh,wgsoh} that the 
experimental data is best supported by the latter scenario, with
the second-neighbor interactions roughly half the nearest-neighbor
interactions.

In a recent paper, Marini and Khomskii\cite{mk}
suggest that the magnetic properties of
the vanadates CaV$_n$O$_{2n+1}$ (with $n=2$, 3, 4)
are determined principally by orbital ordering
and associated structural distortions.  These considerations
lead to rather different spin Hamiltonians than those considered
by most authors previously,\cite{sano,ueda} who neglected the 
orbital degrees of freedom.
To leading order Marini and Khomskii find the system to be a set of
weakly coupled dimers. 

In another recent paper, Pickett\cite{pickett} has used LDA calculations to
study the magnetic properties of CaV$_4$O$_9$.
His most intersting conclusion is that the relevant orbitals are
$d_{x^2-y^2}$  --- different from the ones considered by Marini and
Khomskii. Furthermore, he argues that out-of-plane distortions in
the arrangement of the vanadium atoms are quite important. 
The one-fifth depleted square lattice of vanadium atoms is actually 
made up of two (planar) layers, slightly above and below one 
$ab$-plane, and each forming
a square lattice of ``meta-plaquettes.''
These meta-plaquettes are 
to be distinguished from the plaquettes considered before 
by Ueda {\it et al.}\cite{ueda} and other authors in 
the ``plaquette RVB'' (PRVB) scenario. 
The plaquettes involve nearest-neighbor spins, two
of them lying in the upper layer and two in the lower layer. In this
case, a significant second-neighbor interaction is expected within
the plaquettes. In contrast, the meta-plaquettes of Pickett lie
either entirely in the upper or entirely in the lower layer. 
Pickett argues that although the spins
within a meta-plaquette are further apart than spins between two neighboring
meta-plaquettes, various quantum-chemical arguments conspire to make
the interactions within meta-plaquettes the strongest.
He finds that to leading approximation
the magnetic system should be regarded as decoupled meta-plaquettes.
One important point to note is that within a meta-plaquette the
interactions are expected to be primarily between neighboring spins only.

The aim of this paper is to examine the consistency of the above two
scenarios with existing and prospective experimental data on 
CaV$_4$O$_9$. Neither Marini and
Khomskii nor Pickett provide quantitative estimates for all the 
relevant exchange
constants. Within their scenarios one might 
easily consider a four parameter
Hamiltonian in each case. In the interest of simplicity and because
our main conclusions are unaffected by the number of parameters, we have
considered only a two-parameter Hamiltonian in both cases.

First we consider the proposal of Marini and Khomskii.  They suggest
that the spins in CaV$_4$O$_9$ form antiferromagnetic
dimers with exchange
interaction $J$ approximately equal to the observed triplet
gap, a bit larger than $100\,$K.
If the inter-dimer interactions are
neglected then the susceptibility follows $\chi(T)\propto
T^{-1}/(e^{J/T}+3)$.  This form is shown as the dashed line in 
Fig.~\ref{dimersusc} and compared with
the data of Taniguchi {\it et al.}\cite{tani} (solid line). 
Here we have chosen $J$ and the normalization for the susceptibility so that
the different curves coincide at the maximum.
The two curves are clearly inconsistent.  
Including the ``corner'' exchange constants (shown by thin solid
lines in Fig.~\ref{lattice}), suggested to be the
second most imporant by
Marini and Khomskii, leads to an alternating
spin-chain Hamiltonian
\begin{equation}
{\cal H}=J \sum_i ({\bf S}_{2i}\cdot {\bf S}_{2i+1}+a {\bf S}_{2i+1}
\cdot {\bf S}_{2i+2})\ .
\end{equation}
Here $a=0$ corresponds to the dimer limit. Accurate calculations for
this model are possible by finite-size diagonalization.\cite{DB}
Once again we adjust the exchange constants and the $g$-factors 
to make the theoretical susceptibility coincide with
the experimental data at its peak.
As shown in Fig.~\ref{dimersusc}, increasing $a$ to $0.5$ still
leaves one far from agreement with the data.
It should be noted that much better fits are obtained within 
the PRVB scenario\cite{gwsoh}.
One can trace the inconsistency of the alternating chain model
with the susceptibility data to the
fact that the model's Curie constant is $J(1+a)/4$, which implies
a much larger $J$ value than what is needed to account for the gap.

We now turn to the scenario presented by Pickett. Here the 
leading exchange constant $J$ couples spins
within a meta-plaquette.
The second most significant interaction is that
between nearest-neighbor vanadium spins. These are
the nearest-neighbor plaquette and dimer interactions 
of Ueda {\it et al.} (shown by thick solid and dashed lines in
Fig.~\ref{lattice}), and we take them to be equal
to each other.
They couple spins on neighboring
meta-plaquettes with three terms in the Hamiltonian.
Let these interactions be of strength $\lambda J$. 
The spin-gap for this model as
a function of $\lambda$ is calculated via perturbation theory to be
\begin{eqnarray}
\Delta/J&&= 1 -(2/3) \lambda 
 -0.25347222 \lambda^2  -0.048747749 \lambda^3 \nonumber\\ 
 &&-0.069181496 \lambda^4 -0.030457074 \lambda^5 +O(\lambda^6)\ .
\label{pickettgap}
\end{eqnarray}
$D$log-Pad\'e approximants suggest that this model orders magnetically
at $\lambda_c\approx 0.88$, with a gap exponent approximately $0.7$,
consistent with a $d=3$ Heisenberg universality class. The Curie-Weiss
constant for the model is given by $J(1+3\lambda/2)/2$. Thus
the observed gap and Curie-Weiss constant can both be accounted for with 
$J\approx 220K$ and $\lambda\approx 0.6$. We note that these parameters
will change if interactions between second neighbor metaplaquettes
(that is, metaplaquettes in the same layer) are considered or if
the nearest-neighbor dimer and plaquette interactions are unequal.
However, the quality of the agreement with the high temperature
susceptibility data is likely to be similar to that obtained
within the PRVB scenario. The high temperature susceptibility is determined
by the Curie-Weiss constant and the $g$-factor. Just as in previous
fits within the PRVB scenario, $g$-values smaller than 2 (roughly 1.7)
will be needed to fit the data.

The question remains as to whether one can experimentally distinguish
a weakly coupled meta-plaquette scenario from 
a PRVB phase stabilized by frustration scenario.
One approach is to measure the dispersion of triplet elementary excitations.
In Pickett's model, the interactions between neighboring meta-plaquettes 
are two plaquette couplings and one dimer coupling.  Assuming
those two types of couplings to have equal strength, as in our
derivation of Eq.~(\ref{pickettgap}),
leads to a spectrum in which the gap minimum
occurs at wavevector $(0,0)$. (Note that this corresponds to incipient
two-sublattice order, not incipient ferromagnetism, due to the geometry
of the meta-plaquettes.)  
The dimer couplings would need to be nearly twice as strong as the
plaquette couplings in order for the gap minimum to occur elsewhere
in the Brillouin zone.
In the PRVB scenario considered
earlier the gap minimum most likely lies at an incommensurate wavevector,
or possibly at $(\pi,\pi)$. 
Thus inelastic magnetic neutron
scattering experiments may help to distinguish between these cases.
Frustrating interactions within plaquettes can 
lead to sharply defined singlet excitations,\cite{wgsoh,mila}
which are absent in Pickett's model.
Thus Raman scattering can also help distinguish between the two 
scenarios.

Acknowledgements: We would like to thank Dr.~Daniel Khomskii and 
Dr.~Warren Pickett for communicating their results to us prior
to publication. This work is supported in part by 
the National Science Foundation under Grants No.
DMR-94-57928 (MPG) and DMR-96-16574 (RRPS).

\begin{figure}
\caption{One-fifth depleted square lattice representing vanadium spins
associated with one $ab$-plane of CaV$_4$O$_9$. 
The shaded circles represent vanadium
atoms slightly above the plane and the open circles represent vanadium atoms
slightly below the plane. The thick solid lines represent the elementary
plaquettes of Ueda {\it et al.}  The thin dashed lines
represent the meta-plaquettes of Pickett. In Marini
and Khomskii's picture the 
largest interactions are the ``dimer interactions''
shown here by thick dashed lines and the second largest ones are the ``corner''
interactions, shown by thin solid lines. In Pickett's picture the
largest interactions are the metaplaquette interactions shown by
the thin dashed lines and the second most important interactions are
the dimer and plaquette interactions shown by the thick dashed and
solid lines respectively.}
\label{lattice}
\end{figure}

\begin{figure}
\caption{Uniform susceptibility for CaV$_4$O$_9$ compared with the alternating
spin-chain model with $a=0$ (dimer limit) and $a=0.5$. The 
exchange constant and the normalization for the susceptibility are scaled
to make the curves coincide at their maxima.}
\label{dimersusc}
\end{figure}

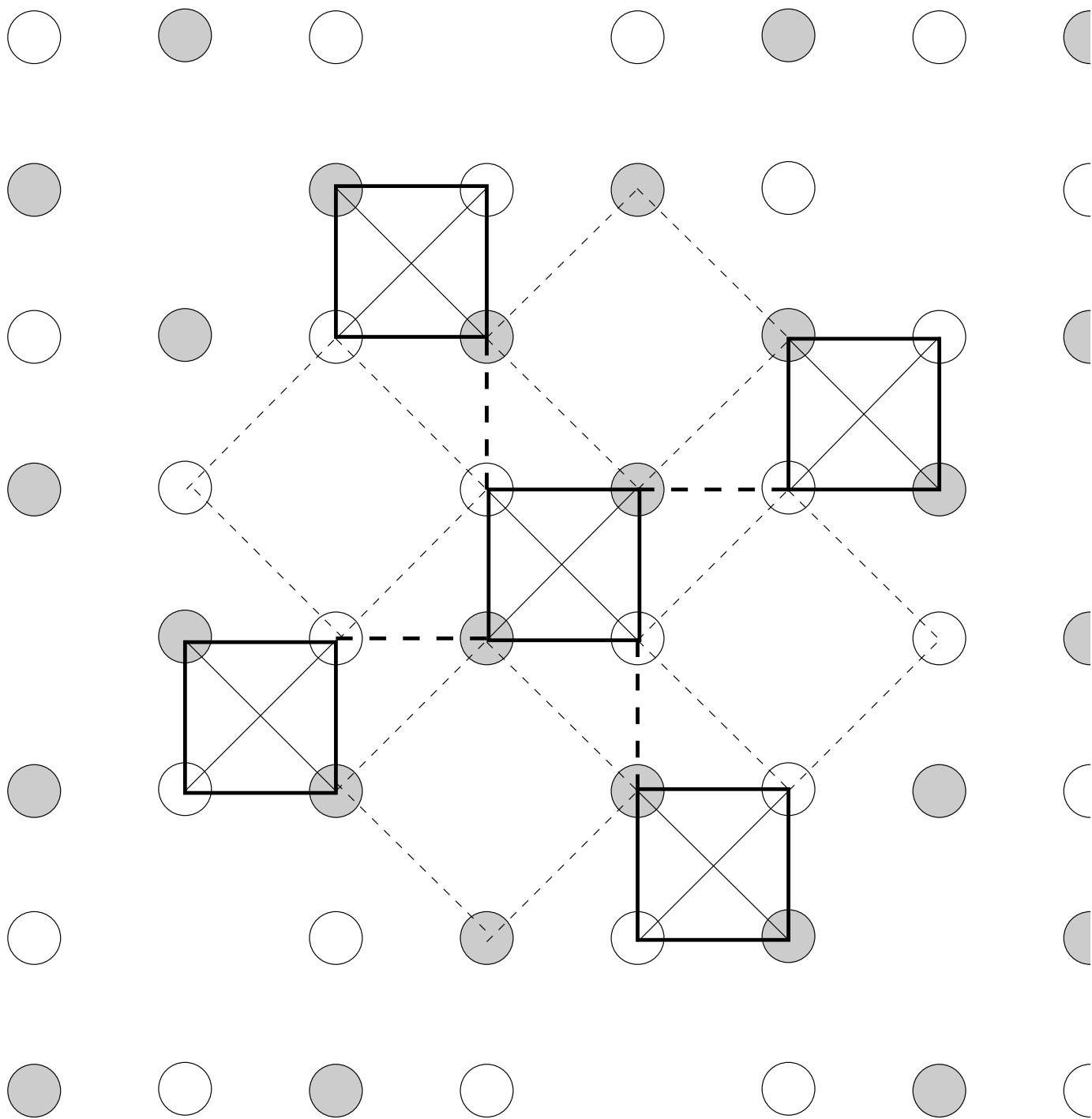
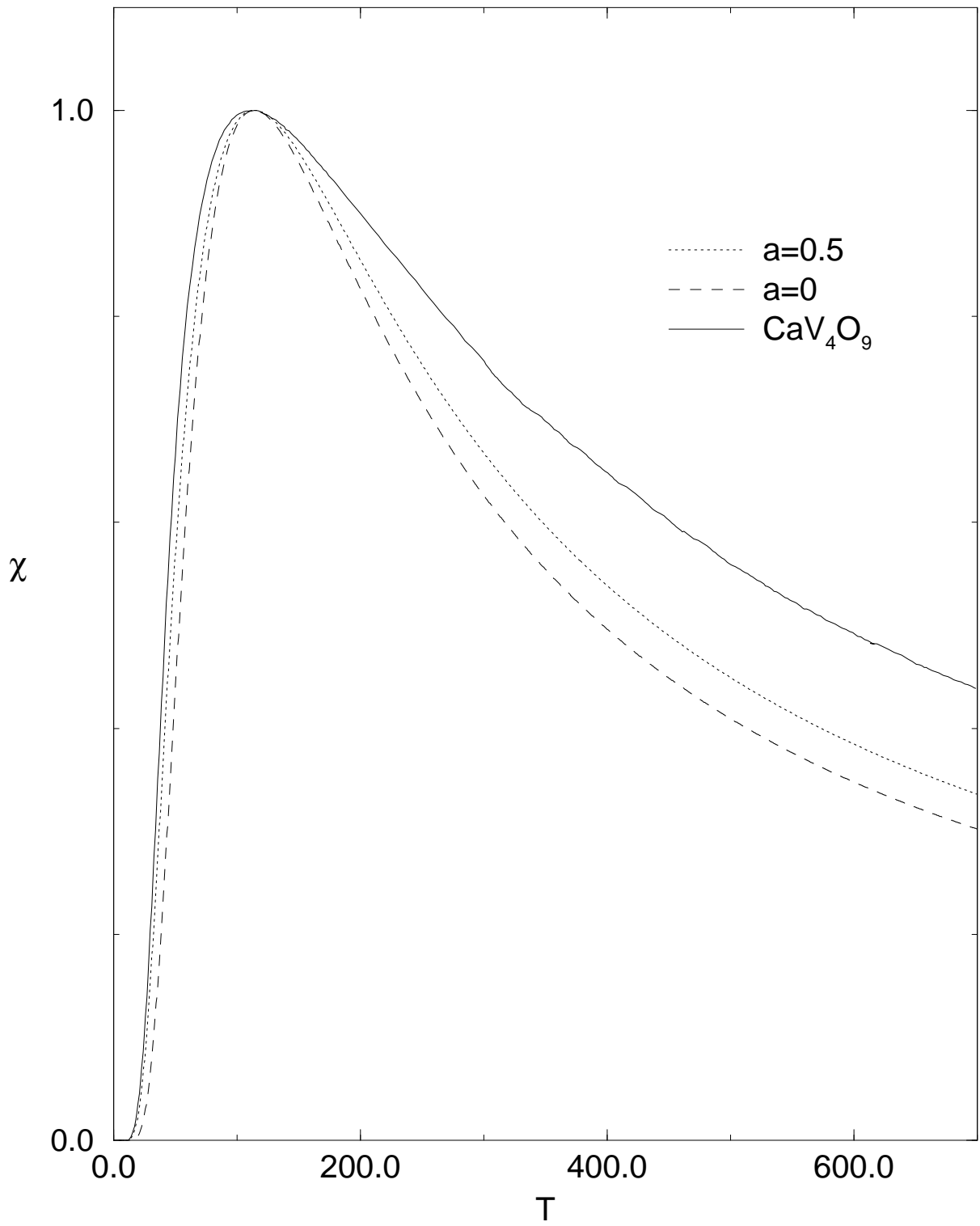
\end{document}